\documentclass[final,3p,times,twocolumn]{elsarticle}

\usepackage{graphicx} 
\usepackage[originalparameters]{ragged2e}
\usepackage{epsfig} 
\usepackage{tikz,pmat}
\usepackage{amsmath} 
\usepackage{amssymb}  
\usepackage{array}
\usetikzlibrary{arrows}
\usetikzlibrary{circuits.ee.IEC}
\usetikzlibrary{shapes.geometric}
\usetikzlibrary{decorations.pathreplacing}
\usetikzlibrary{patterns}
\usepackage{framed}
\usepackage{lineno}
\usepackage{algorithm}
\usepackage{algorithmic}
\usepackage{amsthm}
\usetikzlibrary{decorations.pathmorphing}
\definecolor{ao}{rgb}{0.0, 0.5, 0.0}

\newtheorem{theorem}{Theorem}

\newtheorem{definition}{Definition}
\newcommand{\R}{\mathbb{R}}
\DeclareMathOperator{\conv}{Conv}

\biboptions{compress}

\journal{}

\begin{document}

\begin{frontmatter}

\title{Computing the Approximate Convex Hull in High Dimensions}
\author[CSM1]{Hossein Sartipizadeh \corref{cor1}}
\ead{hsartipi@mines.edu}

\author[CSM1]{Tyrone L. Vincent }
\ead{tvincent@mines.edu}
\address[CSM1]{Electrical Engineering and Computer Science, Colorado School of Mines, Golden, CO 80401}

\cortext[cor1]{Corresponding author: Tel.: +1 7204994943}

\begin{abstract}
In this paper, an effective method with time complexity of $\mathcal{O}(K^{3/2}N^2\log \frac{K}{\epsilon_0})$ is introduced to find an approximation of the convex hull for $N$ points in dimension $n$, where $K$ is close to the number of vertices of the approximation. Since the time complexity is independent of dimension, this method is highly suitable for the data in high dimensions. Utilizing a greedy approach, the proposed method attempts to find the best approximate convex hull for a given number of vertices. The approximate convex hull can be a helpful substitute for the exact convex hull for on-line processes and applications that have a favorable trade off between accuracy and parsimony. 

\end{abstract}


\end{frontmatter}



\section {Introduction}  \label{sec: Intro}

The convex hull of a set of points in $\R^{n}$ is the smallest convex set which contains all the points. This concept is widely used in different fields including data searching and signal processing, clustering, image processing, modeling and robust control \cite{ACC2016Sartipi}, collision detection, and urban planning. 

Many algorithms have been developed to find the convex hull of a set of points.  Some of the first efficient algorithms were Graham's scan method \cite{Graham1972} and the gift wrapping algorithm \cite{chand1970algorithm,Jarvis1973}. For a set of $N$ points these algorithms have $O(N\log N)$ worst case run time for low dimensions ($n=2$ or $3$) which is the same order as the  optimal algorithm. These were followed by many others; a survey can be found in \cite{Ebert2014Survey}. The focus on low dimensions is well motivated by many applications in computational geometry, image processing and computer vision. However, there are also significant applications in higher dimensions.


For higher dimensions, standard methods are the method of  Clarkson and Shor~\cite{Clarkson1989} and the Quickhull algorithm~\cite{Barber1996QuickHull}, which have worst case time complexity in high dimensions of $\mathcal{O}(N^{\lfloor n/2 \rfloor})$. The time complexity of these algorithms are also closely related to the representational complexity of the convex hull. For a set of points, the convex hull is a polytope, and could be represented by the facets, or by the vertices. For a set of $N$ random data in $n$ dimensions, the expected number of vertices is $\mathcal{O}(\log^{n-1}N)$ \cite{Dwyer1988}, thus even in the average case the  representational complexity increases exponentially in the dimension $n$.  
 
This explosion of complexity in higher dimensions can detrimental not only to finding the convex hull, but in using it, as algorithms that use the convex hull as input may require processing each element of the representation. This gives double motivation for working with an {\em approximate} convex hull that covers almost the same space as the true convex hull, but with a reduced representational complexity. First, in order to reduce the computational complexity of finding it, and second the computational burden of using it.  

The search for an approximate convex hull was apparently first considered by Bentley in 1982 \cite{Bentley1982}. This method used sampling within a grid to achieve a worst case time complexity of $\mathcal{O}(N^{\lfloor n/2 \rfloor} + (1/\epsilon)^{n-1}N)$, where $\epsilon$ is the maximum error. An approach that uses a grid over angles in polar coordinates was later proposed by \cite{Xu1998} with a time complexity of $\mathcal{O}((1/\epsilon)^{n-1}N)$. Although providing a trade-off between run time and accuracy, unfortunately in both of these methods also have an exponential dependence on dimension.

In order to tackle convex sets in high dimensions, \cite{Wang2013} introduced a useful norm based method to compute the convex hull, with a stated time complexity of $\mathcal{O}(Nn^{4})$. This method was further extended in \cite{Khosravani2013}. This method could also be modified to determine an approximate convex hull, although it has not been specifically designed for this purpose. Two drawbacks are that this method does not explicitly consider the quality of the approximation as new vertices are found, and that it uses a fairly large number of vertices for initialization. Based on the initialization approach presented in \cite{Wang2013}, this method begins with a $n$-simplex with $n+1$ vertices and therefore is not able to find an optimal approximation with less number of vertices. This would be an issue, for example, in cases when the data closely follows a low dimensional hyperplane in a higher dimension. Moreover, a sparse modeling method for finding the convex hull is introduced in \cite{elhamifar2012} which can also be utilized for finding an approximate convex hull. The time complexity of the method is experimentally about $N^{5}$.

In this work, we present a method for finding an approximate convex hull for which the time complexity is $\mathcal{O}(K^{3/2}N^2 \log K)$  where $K$ is the number of iterations of the proposed algorithm and is close to $V$, the number of vertices of the approximate convex hull, which is usually significantly smaller than $N$. While much slower than existing methods for low dimensions, it is quite feasible to apply even in very high dimensions where  methods with an exponential dependence on dimension would not even be able to run. 

\section{Notation}
$\R$ is the set of real numbers, with $\R^{n}$ a length $n$ vector of real numbers. For $x\in \R^{n}$, $\|x\|$ is the Euclidian norm.  Given a set $\mathcal{S}\subset \R^{n}$, $|\mathcal{S}|$ is the number of elements in $\mathcal{S}$, $\conv{\mathcal{S}}$ is the convex hull of $\mathcal{S}$ and $\mathcal{S}\backslash \mathcal{E}$ is the set obtained by removing the elements in $\mathcal{E}$ from $\mathcal{S}$. Let $\mathcal{P}_{V}(\mathcal{S})$ be the set of all subsets of $\mathcal{S}$ with cardinality $V$ or less.

\section{The Approximate Convex Hull}

Our focus will be to find the vertices of the polytope that will define the convex hull. It is easy to show that these are the extreme points of the set.

\begin{definition} 
Given a finite set $\mathcal{S}=\left\{x_{1},\dots,x_{N}\right\}$ of $N$ unique points in $\R^{n}$, the point $x_{i}$ is an {\em extreme point} of $\mathcal{S}$ if it cannot be represented as a convex combination of points from the set $\mathcal{S} \backslash x_{i}$. 
\end{definition}

The extreme points represent the convex hull in the sense that if $\mathcal{E}$ is the set of extreme points of $\mathcal{S}$ then $\conv{\mathcal{E}} = \conv{\mathcal{S}}$. 

A key basic operation that we will utilize not only to define the approximate convex hull but in our algorithm is to  determine the Euclidean distance of a point $z\in\R^{n}$ to $\conv{\mathcal{S}}$. This function is represented as $d(z,\mathcal{S})$,  and its square can be computed via the quadratic program
\begin{equation}
\begin{aligned}
d(z,\mathcal{S})^{2} & = &\min_{\alpha_{i}} & & &\left\| z - \textstyle\sum_{i=1}^{|\mathcal{S}|}\alpha_{i}x_{i}\right\|^{2} \\ 
& &\mbox{s. t.}& & &\alpha_{i}\geq 0, \quad \textstyle\sum_{i=1}^{|\mathcal{S}|} \alpha_{i} = 1
\end{aligned} \label{eq: QPdistance}
\end{equation}
Note that  $x\in\conv{\mathcal{S}}$ if and only if $d(x,\mathcal{S})=0$.

This QP problem can be solved using the interior point method~\cite{boyd2004convex}. A significant property of the interior point method applied to this problem is that the number of iterations required to reach the optimal solution is independent of $n$, the dimension of $x_{i}$, but rather is dependent on $|\mathcal{S}|$, which is the number of variables. In  \cite{Cai2013} it is shown that the number of iterations of the interior point method for solving a QP with $r$ independent variables is of order $O(\sqrt{r}\log \frac{r}{\epsilon_0})$ where $\epsilon_0$ is the accuracy precision of the interior point method. The complexity of this problem will also be investigated empirically later. 

Since the computational complexity depends on $|\mathcal{S}|$, it is useful for the size of this set to be as small as possible. Fortunately, it is clear that if we have available another set $\mathcal{E}$ (such as the extreme points) such that $\conv{\mathcal{E}} = \conv{\mathcal{
S}}$, then $d(z,\mathcal{E}) = d(z,\mathcal{S})$, and one can use $\mathcal{E}$ for the calculation instead. Finally, we note that the extreme points are the  {\em smallest} subset $\mathcal{E}\subseteq\mathcal{S}$ such that $d(z,\mathcal{E}) = d(z,\mathcal{S})$ for all $z\in\R^{n}$. 

The fact that the extreme points are the smallest subset has a very practical importance in applications that use the convex hull as input, and require computations to be performed on each element of $\mathcal{E}$. Unfortunately, as discussed in the introduction, the number of extreme points tends to greatly increase in high dimension. 

Our objective is to allow for a trade-off between the complexity of the representation of $\conv{\mathcal{S}}$ and its accuracy. A subset $\mathcal{E}\subseteq\mathcal{S}$ would be a completely accurate representation if $d(x,\mathcal{E})=0$ for all $x\in\conv{\mathcal{S}}$. For an approximate representation we will relax the distance to be less than $\epsilon>0$. Specifically, we have the following definition .

\begin{definition}
An {\em $\epsilon$-approximate convex hull} of $\mathcal{S}$ is the convex hull of a minimal subset $\mathcal{E}\subseteq\mathcal{S}$ such that for all $z\in\mathcal{S}$,  $d(z,\mathcal{E})\leq\epsilon$. 
\end{definition}

Note that we are not guaranteed uniqueness, in that there may be multiple $\mathcal{E}$ with the same number of elements, satisfying the distance requirements. However, as $\epsilon$ increases, the size of a minimal representation, $\mathcal{E}$, may be able to decrease.

\section{Finding the Approximate Convex Hull}

The quality of an approximate convex hull with vertices $\mathcal{E}$ is defined by both $\epsilon$ and $|\mathcal{E}|$, the number of vertices.  Thus there is a Pareto front of optimal solutions. To explore this front, one could either minimize $\epsilon$ for a fixed number of vertices or  minimize the number of vertices for a fixed error $\epsilon$; we will focus on the former problem.  We formulate the following optimization problem to find the approximate convex hull: find the subset of $\mathcal{S}$ of cardinality $V$ or less which minimizes worst case distance of a point in $\mathcal{S}$ to the convex hull of the subset, which is notated as
\[
\min_{\mathcal{E} \in \mathcal{P}_{V}(\mathcal{S})} \max_{z\in\mathcal{S}} d(z,\mathcal{E}). 
\]
Unfortunately, this is a combinatorial optimization, requiring a search over each element of $\mathcal{P}_{V}(\mathcal{S})$. In order to achieve an acceptable performance, we will introduce a greedy method that will provide a suboptimal solution. However, this method will still be able to achieve any desired $\epsilon$ performance, although with a potentially larger cardinality than optimal.

\subsection{Basic Greedy Algorithm} \label{sec: BasicGreedyAlgorithm}

The greedy algorithm finds each element of $\mathcal{E}$ sequentially. That is, suppose at step $k$, we have identified a set $\mathcal{E}_{k}$ with $|\mathcal{E}_{k}|=k$. We then find
\begin{equation}\label{eqn:greedystep}
\hat{x} = \arg\min_{x \in \mathcal{S}\backslash\mathcal{E}_{k}} \max_{z\in\mathcal{S}\backslash\mathcal{E}_{k}} d(z,\mathcal{E}_{k}\cup x),
\end{equation}

and set $\mathcal{E}_{k+1} = \mathcal{E}_{k}\cup \hat{x}$. This method initiates with the empty set and terminates either after a fixed number of steps $k$, or after the desired $\epsilon$ has been reached. This optimization problem can be solved by searching over all elements of $\mathcal{S}\backslash\mathcal{E}$ for the outer minimization (as well as over all elements of $\mathcal{S}\backslash\mathcal{E}$ in the inner maximization). While this implies a time complexity of $O(VN^2)$ to reach a cardinality of $V$, this is a vast improvement to searching over $\mathcal{P}_{V}(\mathcal{S})$. In addition, we will introduce methods that can reduce the search space as the iterations progress. 
 
\subsection{Improved Greedy Algorithm} 
 
In this section, two strategies are utilized to speed up the basic greedy method introduced in \ref{sec: BasicGreedyAlgorithm}. First, the solution to \eqref{eqn:greedystep} is explored in more detail and a searching method is introduced to reduce the required time of solving \eqref{eqn:greedystep}. Also, we find and remove elements of $\mathcal{S}$ that are interior points of $\mathcal{E}_{k}$ during processing.  

 \subsubsection{Directed Search} \label{sec: FastSearchingMethod}
 
Let $z_{i}$ and $x_{j}$ be enumerations of  $\mathcal{S}\backslash\mathcal{E}_{k}$. Let $E$ be a matrix with $i,j$th element given by  

\[
E_{i,j} = d(z_{i},\mathcal{E}_{k}\cup x_{j})
\]

The solution to  \eqref{eqn:greedystep} can then be obtained by solving
\[
\hat{j} = \arg \min_{j}\max_{i} E_{i,j}
\]
and setting $\hat{x} = x_{\hat{j}}$. In other words, the maximum value of each column of $E$ is evaluated, and then minimum among those maximums is found.  However, creating $E$ requires $N^2$ evaluations of the distance function $d$.

Improvement can be obtained by searching for the optimal element in a more systematic way. To explain this method, an illustrative example is given. Let $N=4$ and suppose this is the first iteration, and therefore $E$ is a square matrix with 16 elements. Suppose $E$ is populated with the given numbers in Figure \ref{fig: EMatrix}. The maximum of the columns are 6, 7, 4, and 5. The minimum of these is 4 and therefore $\min_{j}\max_{i} E=4$ and $\hat{j} = 3$.      

 \begin{figure}
 \centering
\begin{tikzpicture} [thick,scale=0.9, every node/.style={scale=0.9}]
\fill[gray!30] (0,4) -- (4,4) -- (4,3) -- (0,3);
\fill[gray!30] (1,3) -- (3,3) -- (3,2) -- (1,2);
\fill[gray!30] (2,2) -- (3,2) -- (3,0) -- (2,0);

\draw[step=1cm,gray,thick]
(0,0) grid (4,4);
\draw [-] (-0.5,4.5) node[below] {$i$}  node[right]{$j$}  -- (0,4) ;
\node at (0.5,4.5) {1}; \node at (1.5,4.5) {2}; \node at (2.5,4.5) {3}; \node at (3.5,4.5) {4};
\node at (-0.5,0.5) {4}; \node at (-0.5,1.5) {3}; \node at (-0.5,2.5) {2}; \node  at (-.5,3.5) {1};

\node at (0.5,3.5) {5}; \node at (1.5,3.5) {1}; \node at (2.5,3.5) {2}; \node at (3.5,3.5) {5};
\node at (0.5,2.5) {1}; \node at (1.5,2.5) {5}; \node at (2.5,2.5) {1}; \node at (3.5,2.5) {2};
\node at (0.5,1.5) {0}; \node at (1.5,1.5) {2}; \node at (2.5,1.5) {2}; \node at (3.5,1.5) {5};
\node at (0.5,0.5) {6}; \node at (1.5,0.5) {7}; \node[circle,draw] at (2.5,0.5) {4}; \node at (3.5,0.5) {0};

\end{tikzpicture}
\caption{Illustrative example: Matrix $E$}
\label{fig: EMatrix}
\end{figure}

Instead of computing all elements of $E$, we can calculate only those that are necessary. The key observation is that the maximum of any subset of element in a column provides a lower bound on the maximum over the entire column.   In the given example, suppose only the elements in grey are evaluated. All of the elements of the third column have been evaluated, thus the maximum over this column is known to be 4. For the remaining columns, since elements in each column have been evaluated that are greater than 4, the maximum for these columns must be greater than 4. Thus, it is clear that $\min_{j}\max_{i}E_{i,j}$ must occur for the third column, without having to evaluate the remaining elements.  We use this property to develop the following method:

\begin{enumerate}
\item Calculate the first row of $E$. 
\item Find the minimum value of the first row and select the corresponding column as the potential candidate for $\hat{j}$.
\item Compute the next element of that column and re-evaluate which column has the smallest maximum. Set that  column as the new potential candidate. If two columns have the same maximum, select one at random.
\item Continue until all elements of one of the columns has been found and its maximum is less than or equal to the maximum of the other columns over the computed elements. In this case, $\hat{j}$ is the index of that column and the optimal value $\hat{\epsilon}$ is the maximum value of that column.   
\end{enumerate}   

In the best case this only requires computing $2N-1$ elements, although it can be $N^2$ for the worst case. 

This method is implemented in Algorithm \ref{alg: MinMaxSolution}. This implementation stores the current maximum for each column in the vector $\tilde{E}\in \mathbb{R}^n$, along the row index of currently evaluated elements for each column in vector $\tilde{C} \in \mathbb{R}^n$.  $\tilde{E}$ is set to the first row of $E$ at the beginning and $\tilde{C}$ to vector 1 which indicates that one element of each column has been computed. Then in each step, the next element of the column with the minimum value is computed. If the new element is greater than the previous value, it is saved as the greatest value in $\tilde{E}$. The procedure is continued until the counter shows that we have calculated all elements of a column and its corresponding element in $\tilde{E}$ is still the minimum of $\tilde{E}$. 
\begin{algorithm}[h!]
\begin{algorithmic}[1]
\STATE \bf{Input:} $\mathcal{S}$, $\mathcal{E}$, $N=|\mathcal{S}\backslash\mathcal{E}|$ 
\STATE \textmd{Calculate the first row of $E$.}
\STATE \textmd{Set $\tilde{E}$ to the first row of $E$ and $\tilde{C}$ to $\bf{1}^{1 \times N}$. }
\STATE \textmd{Find the minimum of $\tilde{E}$ and indicate its index with $\hat{j}$.}
\STATE $\tilde{C}_{\hat{j}} = \tilde{C}_{\hat{j}}+1$.
\WHILE{$\max \tilde{C}\leq N$} 
\STATE \textmd{Calculate  $E_{\tilde{C}_{\hat{j}},\hat{j}}$.}
\STATE \textmd{Set $\tilde{E}_{\hat{j}}$ to $\max (\tilde{E}_{\hat{j}}, E_{\tilde{C}_{\hat{j}},\hat{j}})$.}
\STATE \textmd{Find the minimum of $\tilde{E}$ and indicate its index with $\hat{j}$.}
\STATE $\tilde{C}_{\hat{j}} = \tilde{C}_{\hat{j}}+1$.
\ENDWHILE 
\STATE $\hat{\epsilon}=\tilde{E}_{\hat{j}}$ 
\STATE \bf{Output:} $\hat{j}$, $\hat{\epsilon}$
\end{algorithmic}
\caption{Finding min max of a matrix}
\label{alg: MinMaxSolution}
\end{algorithm}

\subsubsection{Interior Points}

The algorithm can be improved by finding and eliminating interior points from consideration.

\begin{definition} \label{def:InteriorPoint}
$z\in\mathcal{S}\backslash \mathcal{E}_{k+1}$ is an {\em interior point} of $\conv \mathcal{E}_{k+1}$ if  $d(z,\mathcal{E}_{k+1})=0$. 
\end{definition}

It can be shown that interior points could be removed from $\mathcal{S}$ as the iterations progress without any changes in the calculated approximate convex hull.

Since $E_{i,j} = d(z_{i},\mathcal{E}_{k}\cup x_{j})$, the interior points of $\mathcal{E}_{k+1}$ can be found via zeros in the $\hat{j}$-th column of $E$. Let $\mathcal{I}_{k}^{\rm tot}$ be the set of interior points of $\mathcal{E}_{k}$, which have been identified at prior steps. At iteration $k$, the directed search is performed over $(\mathcal{S}\backslash\mathcal{I}_{k}^{\rm tot})\backslash\mathcal{E}_{k}$, and from this the interior points of $\mathcal{E}_{k}$ from $\mathcal{S}\backslash\mathcal{I}_{k}^{\rm tot}$ are identified from zeros in the $\hat{j}$-th column, and placed in set $\mathcal{I}_{k}$. The set $\mathcal{I}_{k+1}^{tot}=\mathcal{I}_{k}^{tot} \cup \mathcal{I}_{k}$ is defined, and the next iteration is performed.  This approach significantly decreases the computational time since in each step we remove some of the points for the further processing. It also guarantees that the algorithm acts as preferred and at step $k+1$ will not choose an interior point of $\mathcal{E}_{k+1}$ as the next element of the approximate convex hull.

In addition, it may be that elements of $\mathcal{E}_{k_{0}}$ become interior points of  $\mathcal{E}_{k}$ for some $k>k_{0}$. These points can be removed either at the end of each iteration, or at the completion of the algorithm. This reduces the cardinality of the returned vertex set, which is desirable.

\subsubsection{Initialization}

In the first iteration of the greedy method ($k=1$) \eqref{eqn:greedystep} returns the element closest to the center mass of set $\mathcal{S}$ which is the best approximation when we want to describe set $\mathcal{S}$ only by one element ($V=1$). However, since we aim to cover the space as much as we can, for $V>1$ it is better to start from an extreme point of $\conv\mathcal{S}$. We use the following result to select an extreme point of $\mathcal{S}$ as $\mathcal{E}_{1}$. 

\begin{theorem}
Let $\mathcal{S}$ to be a set of $N$ points in $n$ dimension. Any element of $\mathcal{S}$ which has a {\em minimum} or {\em maximum} in one of the dimensions, is an extreme point of $\conv\mathcal{S}$. 
\end{theorem}

\begin{algorithm}[h!]
\begin{algorithmic}[1]
\STATE \bf{Input:} $\mathcal{S}=\{x_i\}_{i=1}^{N}$, $V$, $\epsilon_{des}$  
\STATE Initialize \textmd{$\mathcal{E}$ and ${\mathcal{S}}'=\mathcal{S}$.}
\WHILE{$|\mathcal{E}|< V$ and $\epsilon>\epsilon_{des}$} 
 \FOR {\textmd{$x_{i} \in {\mathcal{S}}'\backslash\mathcal{E}$ and $z_{j} \in {\mathcal{S}}'\backslash\mathcal{E}$ }}
 \STATE \textmd{Find $\hat{j}$ and $\epsilon$ by Algorithm \ref{alg: MinMaxSolution} where $\hat{j}=\arg \min_{j} \max_{i} d(z_{i},\mathcal{E}\cup x_{j})$.}
  \STATE \textmd{Using $E_{i,\hat{j}}$ calculated in the last step, Find $\mathcal{I}_{k}$ so that if  $d(z_{i},\mathcal{E}\cup x_{\hat{j}})=0$ then $z_{i} \in \mathcal{I}_{k}$.}
 \ENDFOR
\STATE ${\mathcal{S}}'={\mathcal{S}}'\backslash\mathcal{I}_{k}$, $\mathcal{E}=\mathcal{E}\cup x_{\hat{j}}$;
\FOR {\textmd{$c=1$ to $|\mathcal{E}|$ , $c ++$}}
\IF {$d(\mathcal{E}(c),\mathcal{E})=0$}
\STATE $\mathcal{E}\backslash\mathcal{E}(c)$;
\ENDIF
\ENDFOR
 \ENDWHILE 
\STATE \bf{Output:} $\mathcal{E}$, $\epsilon$
   \end{algorithmic}
\caption{Approximate Convex Hull}
\label{alg: ACH}
\end{algorithm}

\section{Summary}    \label{sec: Conclusion}

In this work, an effective algorithm with time complexity of $\mathcal{O}(K^{3/2}N^2\log \frac{K}{\epsilon_0})$ was proposed to compute the approximate convex hull of a data set with $N$ points in $n$ dimension. $K$, the number of iterations of the greedy method, is close to $V$, the number of vertices of the approximate convex hull. According to the time complexity, this method is highly suitable for the data in high dimensions unlike the family of quick-hull methods. The proposed algorithm uses a greedy method to attempt to find the best approximation to the convex hull for a given number of vertices.

\section*{Acknowledgment} 
This work was supported by ONR grant N00014-12-1-0201.

\bibliographystyle{unsrt}
\bibliography{ConvexHullReview}

\begin{thebibliography}{10}

\bibitem{ACC2016Sartipi}
Hossein Sartipizadeh and Tyrone~L. Vincent.
\newblock Uncertainty characterization for robust mpc using an approximate
  convex hull method.
\newblock In {\em American Control Conference}, 2016.

\bibitem{Graham1972}
R.L. Graham.
\newblock An efficient algorithm for determining the convex hull of a finite
  planar set.
\newblock {\em Information Processing Letters}, 1(4):132 -- 133, 1972.

\bibitem{chand1970algorithm}
Donald~R Chand and Sham~S Kapur.
\newblock An algorithm for convex polytopes.
\newblock {\em Journal of the ACM (JACM)}, 17(1):78--86, 1970.

\bibitem{Jarvis1973}
R.A. Jarvis.
\newblock On the identification of the convex hull of a finite set of points in
  the plane.
\newblock {\em Information Processing Letters}, 2(1):18 -- 21, 1973.

\bibitem{Ebert2014Survey}
T.~Ebert, J.~Belz, and O.~Nelles.
\newblock Interpolation and extrapolation: Comparison of definitions and survey
  of algorithms for convex and concave hulls.
\newblock In {\em Computational Intelligence and Data Mining (CIDM), 2014 IEEE
  Symposium on}, pages 310--314, Dec 2014.

\bibitem{Clarkson1989}
Kenneth~L. Clarkson and Peter~W. Shor.
\newblock Applications of random sampling in computational geometry, ii.
\newblock {\em Discrete \& Computational Geometry}, 4(1):387--421, 1989.

\bibitem{Barber1996QuickHull}
C.~Bradford Barber, David~P. Dobkin, and Hannu Huhdanpaa.
\newblock The quickhull algorithm for convex hulls.
\newblock {\em ACM Trans. Math. Softw.}, 22(4):469--483, December 1996.

\bibitem{Dwyer1988}
Rex~Allen Dwyer.
\newblock {\em Average-case Analysis of Algorithms for Convex Hulls and Voronoi
  Diagrams}.
\newblock PhD thesis, Pittsburgh, PA, USA, 1988.
\newblock Order No. GAX88-17713.

\bibitem{Bentley1982}
Jon~Louis Bentley, Franco~P. Preparata, and Mark~G. Faust.
\newblock Approximation algorithms for convex hulls.
\newblock {\em Commun. ACM}, 25(1):64--68, January 1982.

\bibitem{Xu1998}
Zong-Ben Xu, Jiang-She Zhang, and Yiu-Wing Leung.
\newblock An approximate algorithm for computing multidimensional convex hulls.
\newblock {\em Applied Mathematics and Computation}, 94(2‰):193 -- 226, 1998.

\bibitem{Wang2013}
Di~Wang, Hong Qiao, Bo~Zhang, and Min Wang.
\newblock Online support vector machine based on convex hull vertices
  selection.
\newblock {\em Neural Networks and Learning Systems, IEEE Transactions on},
  24(4):593--609, April 2013.

\bibitem{Khosravani2013}
H.R. Khosravani, A.E. Ruano, and P.M. Ferreira.
\newblock A simple algorithm for convex hull determination in high dimensions.
\newblock In {\em Intelligent Signal Processing (WISP), 2013 IEEE 8th
  International Symposium on}, pages 109--114, Sept 2013.

\bibitem{elhamifar2012}
Ehsan Elhamifar, Guillermo Sapiro, and Rene Vidal.
\newblock See all by looking at a few: Sparse modeling for finding
  representative objects.
\newblock In {\em Computer Vision and Pattern Recognition (CVPR), 2012 IEEE
  Conference on}, pages 1600--1607. IEEE, 2012.

\bibitem{boyd2004convex}
Stephen Boyd.
\newblock {\em Convex Optimization}.
\newblock Cambridge University Press, 2004.

\bibitem{Cai2013}
Xinzhong Cai, Guoqiang Wang, and Zihou Zhang.
\newblock Complexity analysis and numerical implementation of primal-dual
  interior-point methods for convex quadratic optimization based on a finite
  barrier.
\newblock {\em Numerical Algorithms}, 62(2):289--306, 2013.

\end{thebibliography}

\end{document}